\documentclass[10pt,twocolumn]{IEEEtran}
\usepackage{amsmath,amssymb,graphicx,epsfig, cite,algorithm,algorithmic, url}
\usepackage[utf8]{inputenc}
\usepackage[mathscr]{euscript}

\newcommand{\y}{\mathbf{y}}

\newcommand{\e}{\mathbf{e}}

\newcommand{\ba}{\mathbf{a}}
\newcommand{\A}{\mathbf{A}}
\newcommand{\bB}{\mathbf{B}}
\newcommand{\bF}{\mathbf{F}}
\newcommand{\w}{\mathbf{w}}
\newcommand{\W}{\mathbf{W}}

\def\minwrt[#1]{\underset{#1}{\text{minimize }}}
\def\maxwrt[#1]{\underset{#1}{\text{max }}}
\def\argminwrt[#1]{\underset{#1}{\text{arg min }}}
\def\argmaxwrt[#1]{\underset{#1}{\text{arg max }}}
\newcommand{\norm}[1]{\left\lVert#1\right\rVert}

\newcommand{\trace}[1]{\text{tr}\left(#1\right)}

\def\bA{{\bf A}}

\def\bF{{\bf F}}

\def\bW{{\bf W}}
\def\bX{{\bf X}}

\def\ba{{\bf a}}

\def\be{{\bf e}}

\def\bt{{\bf t}}

\def\bw{{\bf w}}

\newcommand{\bmu}{\boldsymbol{\mu}}

\newcommand{\btheta}{\boldsymbol{\theta}}

\newcommand{\bOmega}{\boldsymbol{\Omega}}

        \makeatletter
        \def\fps@eqnfloat{!t}
        \def\ftype@eqnfloat{4}
        
        \newenvironment{eqnfloat*}
               {\@dblfloat{eqnfloat}}
               {\end@dblfloat}
        \makeatother
\renewenvironment{thebibliography}[1]{%
 \begin{oldthebibliography}{#1}%
 \setlength{\parskip}{0ex}%
 \setlength{\itemsep}{0ex}%
}%
{%
\end{oldthebibliography}%
}%

\title{Designing Sampling Schemes for Multi-Dimensional Data \thanks{Initial work treating the formulation of this paper has been accepted for publication at the EUSIPCO 2017 and Asilomar 2017 conferences. This work was supported in part by the Swedish Research Council, Carl Trygger's foundations, and the Swedish strategic research program eSSENCE.}}
\author{Johan Sw{\"a}rd$^*$, Filip Elvander$^*$, and Andreas Jakobsson$^*$\thanks{$^*$Dept. of Mathematical Statistics, Lund University, P.O. Box 118, SE-221 00 Lund, Sweden, email: \texttt{\{js,filipelv,aj\}@maths.lth.se}.}}

\begin{document}
\maketitle

\begin{abstract}
In this work, we propose a method for determining a non-uniform sampling scheme for multi-dimensional signals by solving a convex optimization problem reminiscent of the sensor selection problem. The resulting sampling scheme minimizes the sum of the Cram\'er-Rao lower bound for the parameters of interest, given a desired number of sampling points. The proposed framework allows for selecting an arbitrary subset of the parameters detailing the model, as well as weighing the importance of the different parameters. Also presented is a scheme for incorporating any imprecise \emph{a priori} knowledge of the locations of the parameters, as well as defining estimation performance bounds for the parameters of interest. Numerical examples illustrate the efficiency of the proposed scheme.
\end{abstract}

\section{Introduction}
Determining how to suitably sample a signal is an important problem in many signal processing applications, such as sensor positioning and selection in network monitoring \cite{LiuFMV14_62,Jamali-RadSML15_63}, localization and tracking \cite{ChepuriLV13}, magnetic resonance imaging (MRI) \cite{RavishankarB11}, graph signal processing \cite{GamaMMR16_asilomar,AnisGO16_64}, and selecting the temporal sampling \cite{SchmiederSWH93_3}.
In general, these problems can be viewed as sampling a multi-dimensional field containing partly known signal components. For high-dimensional data, it quickly becomes infeasible to sample the field uniformly, especially, in areas such as nuclear magnetic resonance (NMR) spectroscopy when examining living cells, which have limited lifetimes. For example, a recent study of \mbox{4-D} NMR measurements that would have taken about 2.5~years to perform using regular sampling was shown to be possible to construct in merely 90~hours using a non-uniform sampling scheme \cite{KazimierczukZK10_205}. This has caused an interest in formulating sampling schemes for NMR signals, allowing for notable improvements \cite{SchmiederSWH93_3,HybertsTW10_132,Sidebottom16_54,KazimierczukO11_50,KazimierczukO12_223}. 
%
%
%
\par
Among the developed schemes are some exploiting a compressive sensing framework, allowing for an accurate signal reconstruction using fewer samples than the Nyquist-Shannon sampling theorem necessitates for uniformly sampled signals (see, e.g., \cite{KazimierczukO11_50,KazimierczukO12_223,HybertsARW14_241,AotoFKW14_246}). 
However, the developed schemes typically do not optimize the sampling scheme with respect to the expected signals, even though these are often fairly well known. In this work, we strive to exploit this knowledge in order to design a sampling scheme that would allow for a optimal estimation accuracy given the assumed prior knowledge. 

There are many related problems to the here studied sampling scheme problem. In \cite{OzcelikkaleOA10_58}, the problem of how to optimally measure a signal in problems related to propagating wave-fields was studied. More specifically, the authors studied how to best recover the input wave field from noise measurements of the output field given that each measurement is associated with a cost, where the selected cost was set higher for measurement devices with better resolution. The results were presented as trade-off curves between the error of estimation and the total cost budget. In \cite{YilmazLW16_64}, a framework for joint hypothesis testing and estimation  using a minimal sampling size was developed. The proposed framework guarantees, under a Bayesian setup, that the overall detection and estimation performance, given the minimization of the samples size, is the best possible. In \cite{KekatosGW12_27}, the optimal placement of phasor measurement units on power grids was studied.  Other works have been studying problems related to sampling in random fields \cite{GulcuO17_65, ZhangMK09_57} and wireless sensor networks \cite{LiuMV12}. A notable example of the latter category is \cite{LiuMV12}, where the problem of target tracking in wireless sensor networks is studied. The sensors with the most information are found by utilizing a proposed probabilistic sensor management scheme based on the compressed sensing framework. This scheme is decided based on the probability of transmission at each node, found by maximizing the trace of the Fisher information matrix (FIM). Using this approach, sensors with less information can be discarded, implying that fewer sensors need to communicate, thus leading to energy savings.
%
%

%

Lately, for the related problem of optimal sensor placement, there has been several methods proposed in which the combinatorial problem of selecting a subset of sensors is relaxed using convex optimization. In \cite{JoshiB09_57}, the authors consider the case when signal measurements are linear in the unknown parameters and propose a sensor selection scheme based on solving a convex optimization problem inspired by the determinant criterion (D-optimality) of experimental design \cite{Pukelsheim93}. This work was then developed in \cite{Chepuri15,ChepuriL15_63, Jamali-RadSML15_63,LiuCFMLV16_64,ChepuriL15_22}, wherein the authors consider non-linear measurement equations, as well as replacing D-optimality with the average variance criterion (A-optimality) as a performance measure. Specifically, as A-optimality can be interpreted as the sum of the diagonal elements of the Cram\'{e}r-Rao lower bound (CRLB) for the signal parameters, the problem was formulated as to minimize 
the number of required sensors subject to an upper bound on the resulting diagonal sum of the CRLB. 
Assuming that the bound is tight, the method thus finds a sparse set of sensors, i.e., activates a few out of a set of candidate sensors, while keeping
%
the variance of the estimated parameters below  a fixed level. 
\par
In this paper, we expand on this idea, proposing a method for finding a suitable sampling scheme in order to estimate the parameters for signal models where, in general, the signal measurements are non-linear functions of the unknown parameters. By taking the available prior information of the signal into consideration, we propose a sampling scheme that is found by solving a convex optimization problem that guarantees a bound on the worst case CRLB. The sampling pattern is selected via a variable vector, corresponding to the available sample positions, which is penalized using the $\ell_1$-norm, resulting in a sampling scheme that is limited in the number of samples. Furthermore, we reformulate the optimization problem into a semidefinite program (SDP) problem that allows for more flexibility and can be used for adding additional constraints on the optimization.
In general, when estimating a set of parameters, it might be that the scale of the parameters, as well as the accuracy with which they can be estimated, are significantly different. Also, some of the unknown parameters might be of greater interest than the others; again, using NMR as an example, the signal decay is often of more interest than the signal frequencies, the latter often being relatively well known for a given substance, whereas the former measures the sought interactions.
We here propose to use a weighting scheme in order to allow for a relative balancing of the variances of the different parameters, allowing for designing sampling schemes specifically tailored to yield good estimation accuracy for the parameters of interest.
%
%

%
In some applications, one may assume some prior knowledge of the signal of interest, such as, for example, knowledge of the subspace where the signal parameters are to be found. 
%
%
Again using NMR as an illustrative example, the signals of interest consist of decaying modes, being well modeled as a sum of damped sinusoids. These modes are, as noted, often well known in frequency, at least within some reasonably well defined frequency band, whereas the uncertainty of, and the interest in, the signal decays is often more significant. Typically, the problem of interest is thus to specify the damping parameter as accurately as possible using as few samples as possible.
To allow for this case, we herein propose using a gridding of the parameter space in order to guarantee performance within certain bounds, allowing for uncertainty in the parameters.
%
%

This paper is organized as follows. In section \ref{sec:ProblemStatement}, we introduce the problem statement and derive the proposed optimization problem. In Section \ref{sec:Numerical}, we present extensive numerical simulations and results that validates our proposed method. Finally, in Section \ref{sec:Conclusion}, we conclude upon our work.
\section{Problem statement and proposed sampling scheme}\label{sec:ProblemStatement}
Consider a measured signal $y(\bt_n)$, defined on a $D$-dimensional space with $N$ potential $D$-dimensional sampling points, $\bt_n$, $n = 1, 2,\ldots,N$. It is assumed that the probability density function (pdf) of $y(\bt_n)$, here denoted with $p\left(y(\bt_n);\btheta  \right)$, is parametrized by the parameter vector $\btheta \in \mathbb{R}^P$ and that two samples $y(\bt_n)$ and $y(\bt_m)$ are independent if $\bt_n \neq \bt_m$. FIM for sample $y(\bt_n)$ may then be defined as
\begin{align}
	\bF(\bt_n;\btheta) = \mathbb{E}\left\{ \nabla_{\btheta}\log\left(p(y(\bt_n);\btheta)\right)  \nabla^H_{\btheta}\log\left(p(y(\bt_n);\btheta)\right) \right\}
\end{align}
where $\mathbb{E}\left\{\cdot\right\}, \nabla_{\btheta}$, and $(\cdot)^H$ denote the statistical expectation, the gradient with respect to $\btheta$, and the conjugate transpose, respectively.
The here proposed sampling scheme is designed such that it is optimal in the sense of either minimizing the CRLB of the parameters of interest, given that $M$ of the $N$ potential uniform samples are used, or conversely, to minimize the number of samples used given a desired upper bound on the CRLB of the parameters. It is worth noting that as the potential signal samples are assumed to be independent, for any set of samples indices $\bOmega$, it holds that
\begin{align}
	\sum_{n \in \bOmega}\bF(\bt_n;\btheta)
\end{align}
is the corresponding FIM using this sample scheme.
Let the $N$-dimensional vector $\bw$ denote the possible sampling points in the $D$-dimensional sampling space, such that if the $n$:th index, $w_n$, is set to one, this sampling point is used, whereas if it is set to zero, it is not.
Reminiscent of the case of optimal sensor selection, the resulting sampling design problem may then be formulated as (see also \cite{Chepuri15})
\begin{equation} \label{eq:orig_prob}
\begin{aligned}
	&\minwrt[\bw] &&\trace{\left(\sum_{n=1}^N w_n\bF(\bt_n;\btheta)\right)^{-1}} \\
	& \text{subject to} && \norm{\bw}_1 \leq \lambda \\
	&&& w_n \in \{0,1\}  \;,\; n = 1,2,\ldots,N
\end{aligned}
\end{equation}
where $\lambda>0$ and $\text{tr}(\cdot)$ denotes the trace operator. The choice of objective function is related to the so-called A-optimality criterion from design of experiments \cite{Pukelsheim93} as the trace of the inverse FIM corresponds to the sum of the CRLBs of the signal parameters in $\btheta$. Here, the parameter $\lambda$ constitutes an upper bound on the $\ell_1$-norm of the sample selection vector.
%
%
The sampling design scheme \eqref{eq:orig_prob} is not convex due to the restriction that $w_n$, for $n=1,\dots, N$, is defined over a non-convex set. A convex approximation to this problem may be found by relaxing the binary constraint and instead allowing $w_n$ to take any value in the range $[0,1]$ (see, e.g., \cite{ChepuriL15_63}), resulting in
\begin{equation} \label{eq:first_prob}
\begin{aligned}
	&\minwrt[\bw] && \trace{\left(\sum_{n=1}^N w_n\bF(\bt_n;\btheta)\right)^{-1}}\\
	& \text{subject to} &&  \boldsymbol{1}^T\bw\leq \lambda \\
	&&& w_n \in [0,1]  \;,\; n = 1,2,\ldots,N
\end{aligned}
\end{equation}
where $\boldsymbol{1}$ is a vectors of ones with appropriate dimension. It should be noted that we can here replace $||\bw||_1$ with simply $\boldsymbol{1}^T\bw$, since each element in $\bw$ is equal to or greater than zero. Given a solution $\hat{\bw}$ to \eqref{eq:first_prob}, we define the FIM for the corresponding sampling pattern as
\begin{align}\label{eq:threshold}
	\mathcal{I}(\hat{\bw};\btheta) = \sum_{\ell \in \bOmega} \bF(\bt_\ell;\btheta) , \quad \bOmega = \left\{\ell \mid \hat{w}_\ell > \xi \right\}
\end{align}
where $\xi \geq 0$ is a threshold determining whether a sample weight $\hat{w}_\ell$ should be rounded toward one or zero, i.e., whether the sampling point should be included or not. This formulation allows for the minimization of the sum of the CRLBs given an upper bound on the number of samples used. Note that the problem could alternatively be formulated as minimizing the number of sampling points given an upper bound on the sum of the CRLBs.

However, the sampling design in \eqref{eq:first_prob} does not allow for the case when one is primarily interested in a subset of the available parameters. 
Neither does the formulation take into account that the different parameters might have significantly different variances. For example, for a sum of damped sinusoids, the trace constraint in \eqref{eq:first_prob} will clearly be dominated by the CRLB for the amplitudes, as these are orders of magnitude larger than those of the frequencies, and the optimization will therefore put an
%
%
%
%
%
%
emphasis on minimizing the CRLB of the amplitude parameter. 
In order to allow for sampling schemes that put an emphasis on a selection of the parameters of interest, we recently proposed to introduce a weighting matrix, $\A(\btheta)$, acting upon the FIM in \cite{SwardEJ17_eusipco}. Specifically, instead of minimizing the cost function using the FIM, we proposed to perform the minimization using weighted FIMs
\begin{align}
	\tilde{\bF}(\bt_n;\btheta) = \A(\btheta) \bF(\bt_n;\btheta)\A^T(\btheta) \;,
\end{align}
i.e., performing a linear transformation of the variables and minimizing the sum of the CRLBs corresponding to the transformed parameters $\tilde{\btheta} = \A(\btheta) \btheta$.
However, although this formulation allows for shifting emphasis to the parameters of interest, it does not allow for complete disregard of nuisance parameters as $\bA(\btheta)$ has to be definite in order for the matrix inverse to be defined. In order to allow for an arbitrary weighting, we note the following useful identity holds for an invertible matrix $\bB$,
\begin{align}
	\trace{\bB^{-1}} = \sum_{p=1}^P \e_p^T\bB^{-1}\e_p
\end{align}
where $\be_p$ denotes the $p$th canonical basis vector, i.e., a vector with all its elements equal to zero except the $p$th being equal to one. Furthermore, it is noted that for a positive definite matrix $\bB$, a scalar $\mu$, and an arbitrary vector $\ba$, it follows from the Schur complement (see, e.g., \cite{BoydV04}) that
\begin{align}
	\mu - \ba^T\bB^{-1}\ba \geq 0 \iff \begin{bmatrix} \bB & \ba \\ \ba^T & \mu \end{bmatrix} \succeq \mathbf{0}
\end{align}
where $\bX \succeq \mathbf{0}$ indicates that the matrix $\bX$ is positive semi-definite. Thus, it follows that
\begin{align}
	\minwrt[\bB\succ 0] \ba^T\bB^{-1}\ba
\end{align}
and
\begin{equation}
\begin{aligned}
	&\minwrt[\mu, \bB\succ 0] && \mu\\
	& \text{subject to} &&  \begin{bmatrix} \bB & \ba \\ \ba^T & \mu \end{bmatrix} \succeq \mathbf{0}
\end{aligned}
\end{equation}
are minimized by the same matrix $\bB$. Here, $\bB \succ \mathbf{0}$ indicates that the matrix $\bB$ is positive definite.
%
%
%
\begin{figure}[t!]
        \centering
            \includegraphics[width=0.48\textwidth]{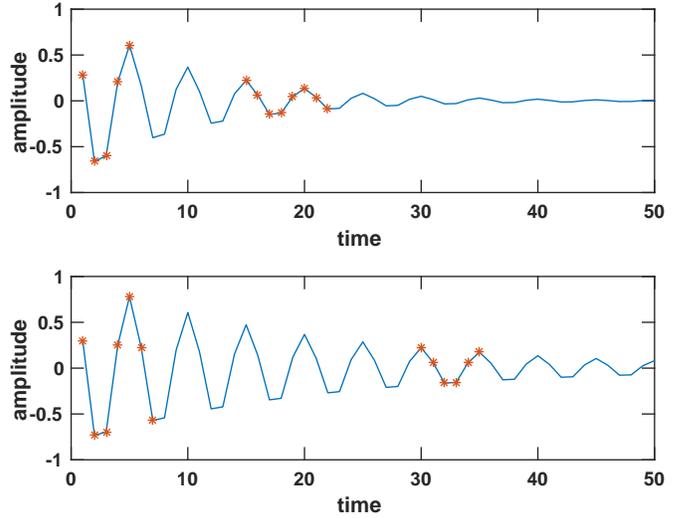}
           \caption{The resulting sample scheme for two different values of $\beta$ plotted against the real part of the signal. The upper most figure details the sampling scheme for $\beta=\frac{1}{10}$ and the bottom figure the sampling scheme for $\beta=\frac{1}{20}$. }
            \label{fig:sampling Example}
\end{figure}
This observation allows us to reformulate \eqref{eq:first_prob} as the semidefinite program (SDP) (cf. \cite{Jamali-RadSML15_63,KekatosGW12_27})
\begin{equation} \label{eq:new_formulation}
\begin{aligned}
	&\minwrt[\bmu,\bw] &&\sum_{p=1}^P \psi_p \mu_p   \\
	& \text{subject to} && \begin{bmatrix} &\sum_{n=1}^N w_n\bF(\bt_n;\btheta)&\be_p\\ &\be_p^T   &\mu_p \end{bmatrix} \succeq \mathbf{0},  \ \forall p \\
	&&& \sum_{n=1}^N w_n\bF(\bt_n;\btheta) \succ \mathbf{0} \\
	&&& \boldsymbol{1}^T\bw\leq \gamma  \quad,\quad w_n\in [0,1], \ \forall n
\end{aligned}
\end{equation}
where $\psi_p$ are weight parameters allowing for putting emphasis on different components of the vector $\btheta$. For example, if $\psi_q = 1$ and $\psi_p = 0$, $\forall p \neq q$, then the CRLB for the parameter $\theta_q$ will be the only one minimized, as $\mu_q$ precisely corresponds to this lower bound, whereas the CRLBs for the other parameters $\theta_p$, $p \neq q$ will be disregarded. Similarly, for $\psi_p = 1$, $\forall p$, the problems \eqref{eq:first_prob} and \eqref{eq:new_formulation} are equivalent.
Another benefit of this formulation is that it allows for a straightforward way of incorporating performance constraints in the minimization problem, such as if, for instance, there is some upper tolerance bound $\lambda_p$ for the CRLB of parameter $\theta_p$. This kind of performance specifications can then be incorporated in the minimization problem via linear inequality constraints according to
\begin{equation}\label{eq:new_formulation_upper_bounds}
\begin{aligned}
	&\minwrt[\bmu,\bw] &&\sum_{p=1}^P \psi_p \mu_p   \\
	& \text{subject to} && \begin{bmatrix} &\sum_{n=1}^N w_n\bF(\bt_n;\btheta)&\be_p\\ &\be_p^T   &\mu_p \end{bmatrix} \succeq \mathbf{0},  \ \forall p \\
	&&& \sum_{n=1}^N w_n\bF(\bt_n;\btheta) \succ \mathbf{0} \\
	&&& \boldsymbol{1}^T\bw\leq \gamma  \quad,\quad w_n\in [0,1], \ \forall n \\\
	&&& \mu_p \leq \lambda_p\:, \ \forall p
\end{aligned}
\end{equation}
Furthermore, one may not only be interested in designing a sampling scheme for a single parameter vector $\btheta$, but rather for a set of parameter vectors.
For example, consider the case when the parameters in $\btheta$ are only partly known, such that one may assume that $\btheta$ instead lies in a set of possible parameters, $\Theta$. In such cases, it may be desired to treat some of the parameters as known, whereas others are only partly known, within some set of uncertainty. To allow for this, as well as taking the weighting into account, we further generalize \eqref{eq:new_formulation_upper_bounds} such that the sampling scheme is designed as
\begin{equation} \label{eq:gridded_prob}
\begin{aligned}
	&\minwrt[\bmu,\bw] &&\sum_{p=1}^P \psi_p \mu_p   \\
	& \text{subject to} && \begin{bmatrix} &\sum_{n=1}^N w_n\bF(\bt_n;\btheta)&\be_p\\ &\be_p^T   &\mu_p \end{bmatrix} \succeq \mathbf{0},  \ \forall p,\forall \btheta \in \Theta  \\
	&&& \sum_{n=1}^N w_n\bF(\bt_n;\btheta) \succ \mathbf{0} \\
	&&& \boldsymbol{1}^T\bw\leq \gamma  \quad,\quad w_n\in [0,1], \ \forall n \\\
	&&& \mu_p \leq \lambda_p\:, \ \forall p
\end{aligned}
\end{equation}
Using this formulation, the optimal $\mu_p$ will, assuming that $\psi_p >0$, now correspond to a worst case CRLB for the $p$th component of $\btheta$, when $\btheta \in \Theta$, i.e., for the obtained sampling sampling scheme
\begin{align}
	\mu_p = \argmaxwrt[\btheta \in \Theta] \be_p^T\mathcal{I}(\hat{\bw};\btheta)^{-1}\be_p
\end{align}
Thus, the solution to \eqref{eq:gridded_prob} is a sampling scheme minimizing the worst case CRLB for the parameters of interest if the parameter vector $\btheta$ is known to be in the set $\Theta$.

Further, one could also consider the case where there is some cost associated with changing sampling points in one of the dimensions. For instance, if one of the sampling dimensions corresponds to a certain setting of a machine, e.g., time delay or magnetic flow, it could be more costly to acquire many different sample points in this dimension. Illustrating this in the \mbox{2-D} case, one could include such a cost in the optimization by forming the $N_1\times N_2$ matrix $\W$ by reshaping the vector $\w$, and adding the constraints
\begin{align}
\left|\left|\bW^T\right| \right|_{2,1}=\sum_{n=1}^{N_1}\left|\left|\bW_{(:,n)}\right| \right|_2\leq \gamma_1\\
\left|\left|\bW\right| \right|_{2,1}=\sum_{n=1}^{N_2}\left|\left|\bW_{(n,:)}\right| \right|_2\leq \gamma_2
\end{align}
to \eqref{eq:gridded_prob}. Here, $\gamma_1$ and $\gamma_2$  are tuning parameters that may be set according to the associated cost. This constraint can easily be omitted simply by setting $\gamma_1=\gamma_2=\infty$.

\begin{figure}[t!]
        \centering
        \hspace*{-5 mm}
            \includegraphics[width=.56\textwidth]{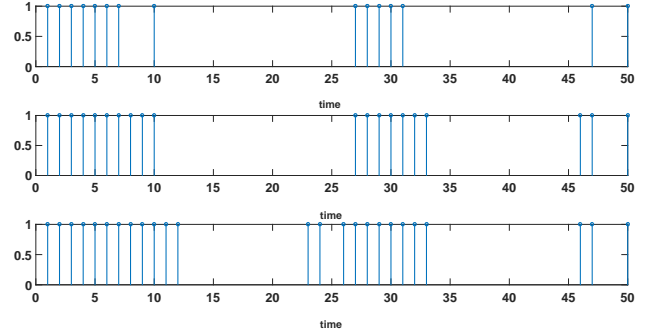}
           \caption{The resulting sample scheme for three different settings of $\gamma$, namely $\gamma=15$, $\gamma= 20,$ and $\gamma=25$, where the signal contains two linear chirps. }
            \label{fig:chirps}
\end{figure}

It is also worth noting that when relaxing \eqref{eq:orig_prob} in favor for \eqref{eq:first_prob}, we can no longer guarantee that the weights are exactly $0$ or $1$. In this case, as is noted in \eqref{eq:threshold}, we simple choose an appropriate threshold such that values above the threshold are deemed as ones, and the values below are deemed as zeros. However, a better approximation of \eqref{eq:orig_prob} is found by using re-weighting. This may be done by first solving \eqref{eq:gridded_prob}, yielding the estimated $\w^{(1)}$, where the superscript $(\cdot)^{(j)}$ denotes $j$th iteration. Then, \eqref{eq:gridded_prob} is solved again, but this time with
\begin{align}
\frac{1}{w_n^{(1)}+\epsilon}
\end{align}
as a scaling factor for each $w_n$, where $\epsilon$ is a small number added to the denominator to avoid numerical problems. This procedure can then repeated until convergence. The re-weighting is a better approximation of the $\ell_0$-norm, and thus is more likely to produce weights with values close to zero or one. As we have empirically found that using re-weighting for the here studied examples offers only a marginal improvement, while significantly increasing the computational cost due to the iterative procedure, we have in our examples chosen to use the simpler thresholding approach. 
%
\section{Numerical results}\label{sec:Numerical}
\subsection{Illustration in 1-D}
To illustrate the proposed sampling scheme, we consider the NMR signal model, as noted being formed as a sum of damped sinusoids (for ease of notation, we initially focus on the \mbox{1-D} case), such that
%
%
\begin{align} \label{eq:model_eq}
	y(t_n) = \sum_{k=1}^{K}\alpha_k \exp\{2i\pi f_k t_n-\beta_k t_n+i\phi_k \} + \epsilon(t_n)
\end{align}
for $n=1,\dots,N$, where $\alpha_k, f_k, \beta_k$, and $\phi_k$ are the frequency, damping, and phase of the $k$:th component, respectively, and where $\epsilon$ is an additive noise term, here assumed to be well modeled as a white, circularly symmetric Gaussian noise with variance $\sigma^2$, with $t_n$ being the time at sample $n$. For simplicity, we consider uniformly sampled candidate sampling times, $t_n$.
As an illustration, Figure~\ref{fig:sampling Example} shows an example of sampling schemes found by solving \eqref{eq:gridded_prob} for two different levels of decay for a single damped sinusoid such that $\beta=1/10$ for the top figure, and $\beta=1/20$ for the bottom figure, but otherwise identical signal parameters. In both cases, $\gamma = 13$ so that $M = 13$ sample points, out of $N = 50$ possible candidates, are selected. Also, $\psi_p = 1$, $p = 1,\ldots,4$, i.e., all signal parameters are considered in the minimization. As can be seen, the placing of the samples are determined by the damping parameter. As may be expected, for both values of $\beta$, some samples are placed in the beginning of the signal, where the signal to noise ratio (SNR) is at its maximum.
To allow for an accurate estimation of the damping constant, one can also note that a further set of samples are selected later in the signal, with the more strongly decaying signal selecting them earlier than the less damped version, agreeing with the intuition that the more rapidly decaying signal contains less information at later sampling times.

As a further example, we next consider an example showing the resulting sample scheme for a signal containing two linear chirp components on the form
\begin{align}
y(t_n)=\sum_{k=1}^2 \alpha_k \exp\left\{2i\pi\left(f^0_k+ f^1_k t_n\right)t_n +i\phi_k\right\}+\epsilon(t_n)
\end{align}
where $f^0_k$ and $f^1_k$ denote the frequency starting point and the slope of the chirp component $k$, respectively. Figure \ref{fig:chirps} shows the three sampling schemes yielded by the proposed method for three different setting on $\gamma$, namely $\gamma=15$, $\gamma=20$, and $\gamma=25$. The here used parameters had the values $\alpha_1=\alpha_2=5$, $f_1^0=0.1$, $f_2^0=0.5$, $f_1^1=0.01$, $f_2^1=-0.003$, and the phases were set to $\phi_1=\pi/2$, and $\phi_2=\pi/3$. Due to the linear drift in frequency, it is reasonable to assume that the resulting sample scheme should have at least two clusters; one in the beginning of the signal, and one at the end of the signal. Looking at the sampling schemes in Figure \ref{fig:chirps} supports this intuition; three clusters are present for all three settings of $\gamma$. When $\gamma$ increases the two first clusters gets bigger, whereas the last cluster remains more or less unchanged.
\begin{figure}[t!]
        \centering
            \includegraphics[width=0.48\textwidth]{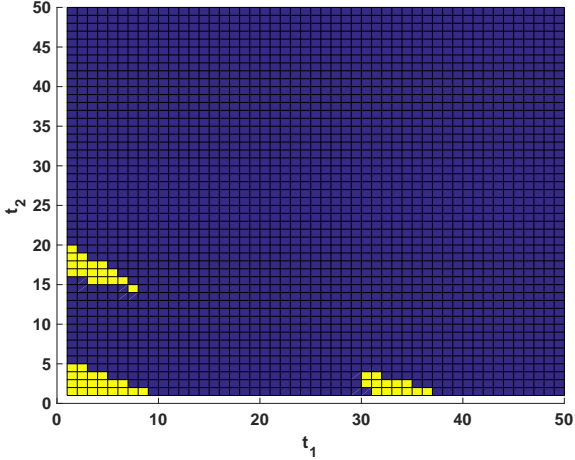}
           \caption{The resulting sampling scheme consisting of 50 selected samples for a signal consisting of a \mbox{2-D} damped sinusoid as found when solving \eqref{eq:new_formulation} with all $\psi_p = 1$.}
            \label{fig:illustration_2D_all_params}
\end{figure}

\subsection{Illustration in 2-D}
As further illustration of the impact of the choice of weight parameters $\psi_p$, consider the \mbox{2-D} case with one damped sinusoid, i.e.,
\begin{align} \label{eq:one_2D_sine}
	y(t_1,t_2) = \alpha \mathrm{e}^{2i\pi (f_1t_1 +f_2t_2) - (\beta_1t_1 + \beta_2t_2)+i\phi} + \epsilon(t_1,t_2)
\end{align}
with $\alpha = 1$, $f_1 = 0.2$, $f_2 = 0.5$, $\beta_1 = 1/20$, $\beta_2 = 1/10$, $\phi = 1/2$, and noise variance $\sigma^2 = 0.1$. Figure~\ref{fig:illustration_2D_all_params} presents the sampling scheme found by solving  \eqref{eq:new_formulation} with $\gamma = 50$, i.e., 50 sampling points are chosen, for the case when $\psi_p = 1$ for all parameters.

As can be seen, the optimal sampling pattern here consists of three clusters of selected sampling points; one close to the origin and two close to the two time axes. Note that this is analogous to the \mbox{1-D} case as the sampling cluster close to the first time axis is located further from the origin due to the decay in the first dimension being slower.

\begin{figure}[t!]
        \centering
            \includegraphics[width=0.48\textwidth]{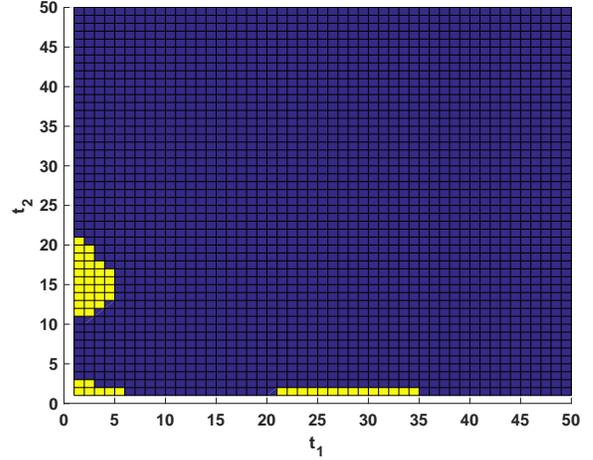}
           \caption{The resulting sampling scheme consisting of 50 selected samples for a signal consisting of a \mbox{2-D} damped sinusoid as found when solving \eqref{eq:new_formulation} with all $\psi_p = 1$ except for the amplitude and phase parameters, for which $\psi_p = 0$.}
            \label{fig:illustration_2D_only_freq_and_damp}
\end{figure}

In contrast, Figure~\ref{fig:illustration_2D_only_freq_and_damp} displays the corresponding scheme found when solving \eqref{eq:new_formulation}, again with $\gamma = 50$, but only giving weight to the frequency and damping parameters, i.e., the $\psi_p$ corresponding to the amplitude and phase parameters are set to zero. As can be seen, assigning the amplitude and phase parameters zero weight has the effect of shifting sampling points away from the origin to the clusters close to the $t_1$ and $t_2$ axes, in order to put more emphasis on the frequency and damping parameters. Indeed, the sum of the CRLBs for the parameters, as given by the sampling scheme in Figure~\ref{fig:illustration_2D_all_params}, is $2.31 \cdot 10^{-2}$, whereas it is $3.61 \cdot 10^{-2}$ for the sampling scheme in Figure~\ref{fig:illustration_2D_only_freq_and_damp}. However, if one considers the sum of the CRLBs for the frequency and damping parameters, these are $6.53 \cdot 10^{-4}$ and $4.42 \cdot 10^{-4}$ for Figures~\ref{fig:illustration_2D_all_params} and \ref{fig:illustration_2D_only_freq_and_damp}, respectively.

%

%
%
\subsection{Simulations in 1-D}
\subsubsection{Optimization vs simulation}
\begin{figure}[t!]
        \centering
            \includegraphics[width=0.48\textwidth]{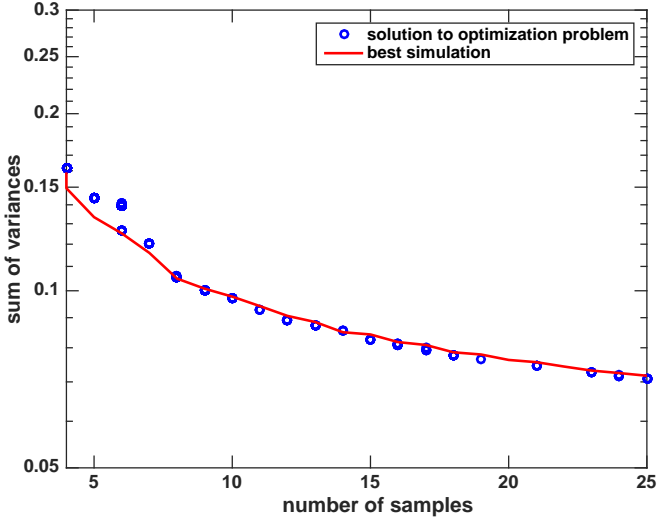}
           \caption{Sum of CRLBs for the parameters, i.e., $\text{tr}\left( \mathcal{I}(\hat{\bw};\btheta)^{-1}  \right)$, for the sampling patterns given by the optimization problem and the best simulation, respectively, for different number of sampling points.}
            \label{fig:opt_vs_sim_noWeight}
\end{figure}

In Figure~\ref{fig:opt_vs_sim_noWeight}, we motivate that solving \eqref{eq:gridded_prob} is indeed a reasonable approach to determine optimal sampling patterns. The figure shows the obtained sum of the CRLBs for the parameters, i.e., $\text{tr}\left( \mathcal{I}(\hat{\bw};\btheta)^{-1}  \right)$, where the sampling pattern is obtained by solving \eqref{eq:gridded_prob} for the case of $K = 1$ using the model \eqref{eq:model_eq}, for a singleton set $\Theta$. This is done for varying values of $\gamma$ such that the number of samples used vary between $M = 5$ and $M = 25$. As a comparison, for each sample size $M$, we carry out $10^6$ Monte Carlo simulations, in which we randomly decide on which $M$ sampling points to use. We then compute which of these $10^6$ sampling patterns that results in the lowest sum of CRLBs. 
As can be seen from the figure, the randomized approach achieves better results for small sample sizes, this as the simulations then become an exhaustive search, i.e., the simulations will with high likelihood find the exact solution to \eqref{eq:orig_prob}. However, as the sample size increases, so does the number of possible sampling patterns, which is $N \choose M$. As can be seen from the figure, the sampling scheme determined by \eqref{eq:gridded_prob} is then able to achieve an optimal performance as the sample size increases.
%
%
%
%
\begin{figure}[t!]
        \centering
        \vspace{-2.05 mm}
            \includegraphics[width=0.48\textwidth]{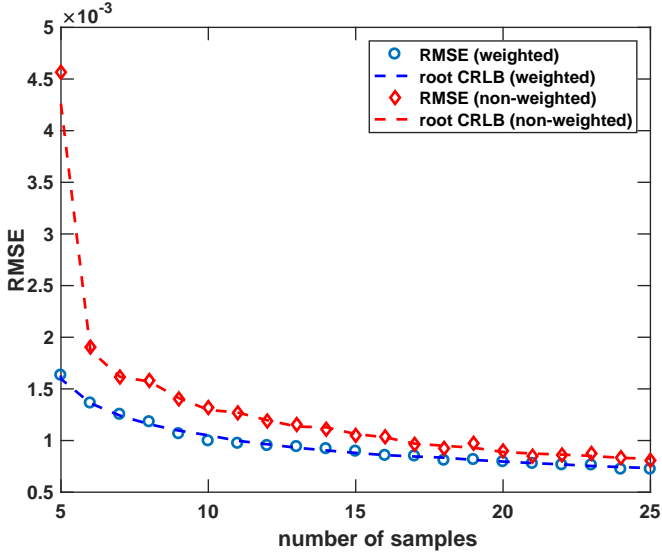}
           \caption{Obtained RMSE for the frequencies, when using the sampling patterns for the weighted and non-weighted cases, respectively.}
            \label{fig:w_vs_nw_freq}
\end{figure}
%
%
\subsubsection{Weighting}
\begin{figure}[t!]
        \centering
            \includegraphics[width=0.5\textwidth]{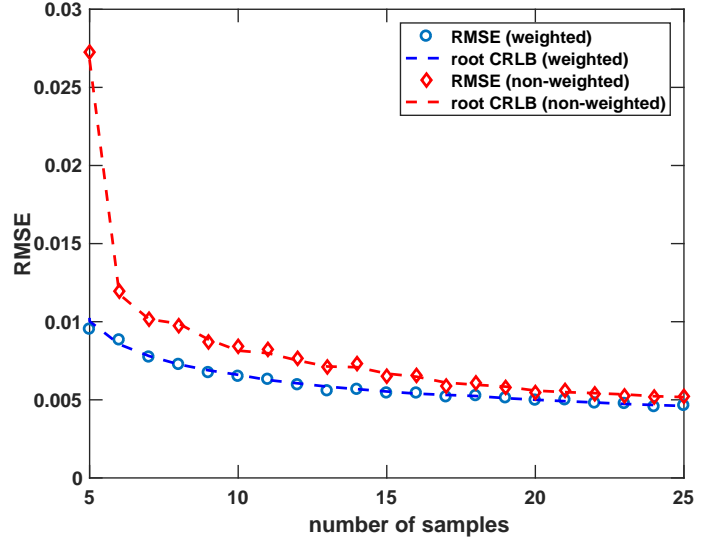}
           \caption{Obtained RMSE for the damping, when using the sampling patterns for the weighted and non-weighted cases, respectively.}
            \label{fig:w_vs_nw_beta}
\end{figure}

In Figures~\ref{fig:w_vs_nw_freq} and \ref{fig:w_vs_nw_beta}, we proceed to examine the effect of using the weighted FIM in \eqref{eq:gridded_prob}.
%
This is done for a signal consisting of two damped sinusoids with parameters $(\alpha_1,f_1,\beta_1,\varphi_1) = (1,0.2,1/12,0.5)$ and \mbox{$(\alpha_2,f_2,\beta_2,\varphi_2) = (1,0.65,1/20,\pi/5)$}. The noise variance was $\sigma^2 = 0.01$ and $N = 50$.
Assuming that we are interested only in the frequencies $f_1, f_2$, and the damping factors $\beta_1, \beta_2$, but not in the amplitudes or the phases, the weight parameters $\psi_p$ are set to one for the frequency and damping parameters, whereas they are set to zero for the amplitudes and phases. Thus, the sought sampling pattern will be designed to increase the accuracy for the frequency and damping  parameters at the expense of the amplitude and phase parameters.

%
%
%
%
%
The resulting root CRLB, as a function of the number of samples used, for the frequencies $f_1$ and $f_2$ and the dampings $\beta_1$ and $\beta_2$ are shown in Figures~\ref{fig:w_vs_nw_freq} and \ref{fig:w_vs_nw_beta}, respectively. The root CRLB for the frequencies $f_1$ and $f_2$ is here defined as the root of the sum the individual CRLBs, and correspondingly for the dampings, $\beta_1$ and $\beta_2$. For comparison, the figures also present the root CRLBs corresponding to the optimal sampling patterns obtained for the case when no weighting is applied, i.e., $\psi_p = 1$, $\forall p$. As can be seen, the weighting scheme results in sampling patterns that decreases the CRLB for the parameters of interest, in this case the frequencies and dampings.
Also plotted is the obtained root mean squared error (RMSE) for the frequency and damping parameters, respectively, obtained when estimating these parameters using non-linear least squares (NLS) applied to simulated signals. The NLS estimate is found by solving
\begin{align}\label{eq:NLS}
\hat{\btheta} = \underset{\btheta}{\text{argmin}}\quad \frac{1}{2}||\y-g(\btheta)||_2^2
\end{align}
where $\y$ is the data and $g(\theta)$ is the (non-linear) data model with parameter $\theta$. In this paper, a minimum of \eqref{eq:NLS} is found by evaluating the cost function over a grid of parameter values $\theta$. The $\theta$ that achieves the lowest value of \eqref{eq:NLS} then becomes the resulting estimates.
  The RMSE is here defined as the root of the sum of the individual MSEs for the frequencies and dampings, respectively.  As can be seen, the RMSE coincides with the root CRLB, implying that the bound is tight.
%
%
%
%
%
%
\subsubsection{Gridding}
\begin{figure}[t!]
        \centering
        \vspace{-2.1 mm}
            \includegraphics[width=0.48\textwidth]{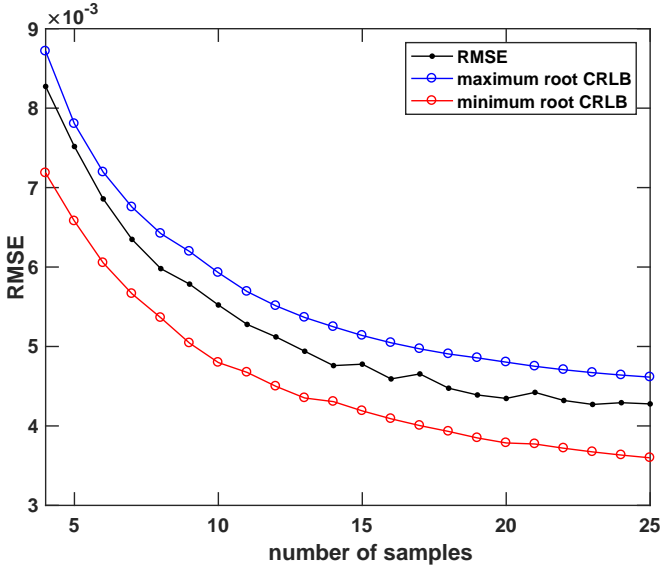}
           \caption{Obtained RMSE for the frequency $f$, when estimating $\btheta$ for the sampling pattern obtained for a grid of damping parameters $\beta$.}
            \label{fig:betaGrid_freq}
\end{figure}

\begin{figure}[t!]
        \centering
            \includegraphics[width=0.48\textwidth]{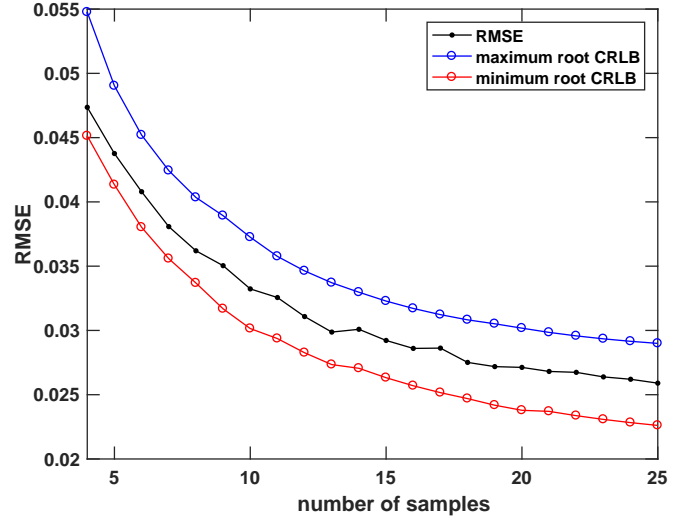}
           \caption{Obtained RMSE for the damping $\beta$, when estimating $\btheta$ for the sampling pattern obtained for a grid of damping parameters $\beta$.}
            \label{fig:betaGrid_damp}
\end{figure}

Figures~\ref{fig:betaGrid_freq} and \ref{fig:betaGrid_damp} show the effect of finding an optimal sampling pattern for a set of parameters $\btheta \in \Theta$ when solving \eqref{eq:gridded_prob}. The results are obtained for a single decaying sinusoid.
Here, we let $\Theta = \left\{\btheta_\ell  \right\}_{\ell=1}^L$ express uncertainty in only the damping parameter $\beta$ by fixing $\alpha, f$, and $\varphi$ and letting $\Theta$  be a gridding over the damping parameter $\beta$, such that the parameter vectors constituting $\Theta$ are $\btheta_\ell = (\alpha,f,\beta_\ell,\varphi)^T$, where
\begin{align} \label{eq:betaGrid}
	\beta_\ell = \beta_\text{lower} + \frac{\ell-1}{L}\Delta_\beta
\end{align}
with $\Delta_\beta$ denoting the grid spacing, in effect letting $\beta$ reside in the uncertainty interval
\begin{align}
	\mathcal{J}_\beta = \left[\beta_\text{lower},\beta_\text{lower}+\frac{L-1}{L}\Delta_\beta  \right]
\end{align}
The parameters used are $\alpha = 1$, $\varphi =0.5$, $\sigma^2 = 0.1$, \mbox{$\beta_\text{lower} = 0.1$}, $\Delta_\beta = 0.022$, and $L = 10$. Using this, we solve \eqref{eq:gridded_prob} to get optimal sampling patterns as the number of samples grows.
To evaluate the performance of the obtained sampling schemes, we then randomly sample the parameter vectors $\btheta$ where $\beta$ is sampled uniformly on $\mathcal{J}_\beta$, i.e., on the interval covered by the grid $\Theta$, but not on the grid points $\beta_\ell$, $\ell =0,1,\ldots,L-1$.
We then estimate $\btheta$ using NLS and compute the RMSE for the parameters $\btheta$. The figures show the obtained MSE using 5000 Monte Carlo simulations for the frequency $f$ and the damping $\beta$, respectively. Also presented are the best and worst case root CRLBs found on the grid $\Theta$ for each parameter. The obtained RMSE lies between the lowest and highest on-grid root CRLB for both parameters and for all considered sample sizes, suggesting that \eqref{eq:gridded_prob} indeed yields sampling schemes with a guaranteed worst case performance, as well as a lower limit on the possible RMSE.
%

%

\subsection{Simulations in 2-D}
\subsubsection{Optimization vs simulation}
\begin{figure}[t!]
        \centering
            \includegraphics[width=0.5\textwidth]{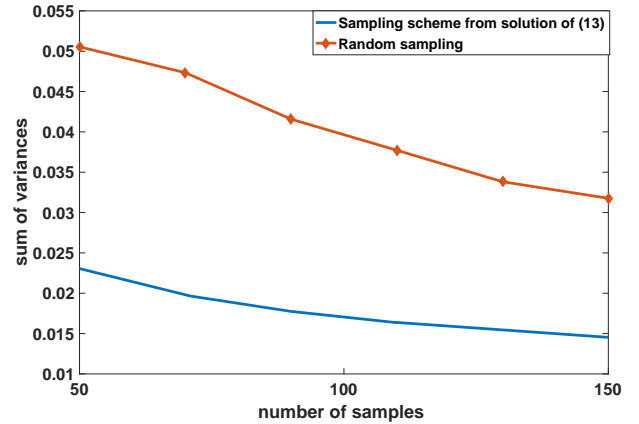}
           \caption{The sum of variances of the parameters of interest as a function of the number of selected samples.}
            \label{fig:crbCompare}
\end{figure}

As was seen in the \mbox{1-D} setting, the optimization scheme was able to outperform the method of randomly selecting sampling points and then choosing the scheme minimizing the sum of the CRLB. In \mbox{2-D}, this becomes even more apparent as the number of potential sampling points increase rapidly with increasing dimension. An illustration of this is shown in Figure \ref{fig:crbCompare}, showing the sum of the CRLBs obtained when solving for varying number of desired sampling points. The signal considered is the \mbox{2-D} damped sinusoid in \eqref{eq:one_2D_sine} with parameters $\alpha = 1$, $f_1 = 0.2$, $f_2 = 0.5$, $\beta_1 = 1/20$, $\beta_2 = 1/10$, $\phi = 1/2$, and $\sigma^2 = 0.1$. We here let $\psi_p = 1$, $\forall p$, and consider a sampling space of $50\times 50$ potential sampling times. Also presented is the sum of the CRLBs for the best (defined as the one with smallest sum of CRLBs) among $10^7$ sampling scheme obtained by randomly choosing sampling points. As can be seen from the figure, the proposed method outperforms the random sampling for all numbers of selected samples. It is worth noting that the computational time to evaluate the $10^7$ sampling schemes was three times longer than solving the proposed problem using a off-the-shelf convex solver \cite{CVX}.
%
%
\subsubsection{Weighting}
\begin{figure}[t!]
        \centering
        \vspace{-2.05 mm}
            \includegraphics[width=0.48\textwidth]{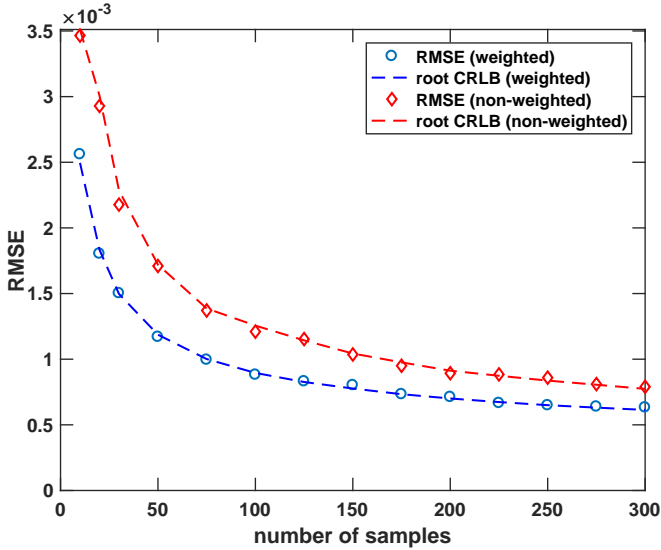}
           \caption{Obtained RMSE for the frequencies in the first dimension, when using the sampling patterns for the weighted and non-weighted cases, respectively.}
            \label{fig:2D_sum_of_freqs_first_dim}
\end{figure}

\begin{figure}[t!]
        \centering
        \vspace{-2.05 mm}
            \includegraphics[width=0.48\textwidth]{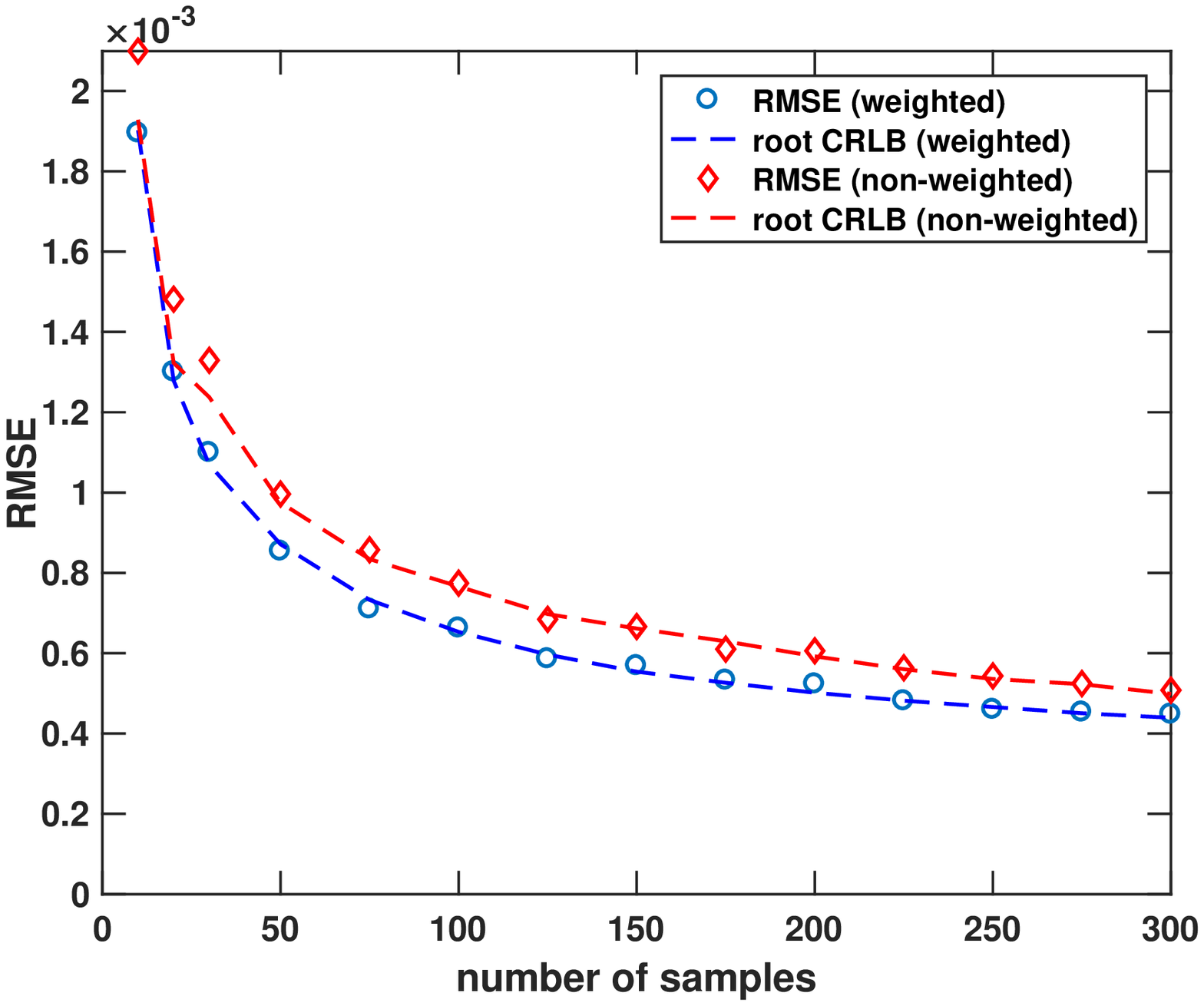}
           \caption{Obtained RMSE for the frequencies in the second dimension, when using the sampling patterns for the weighted and non-weighted cases, respectively.}
            \label{fig:2D_sum_of_freqs_second_dim}
\end{figure}

\begin{figure}[t!]
        \centering
        \vspace{-2.05 mm}
            \includegraphics[width=0.48\textwidth]{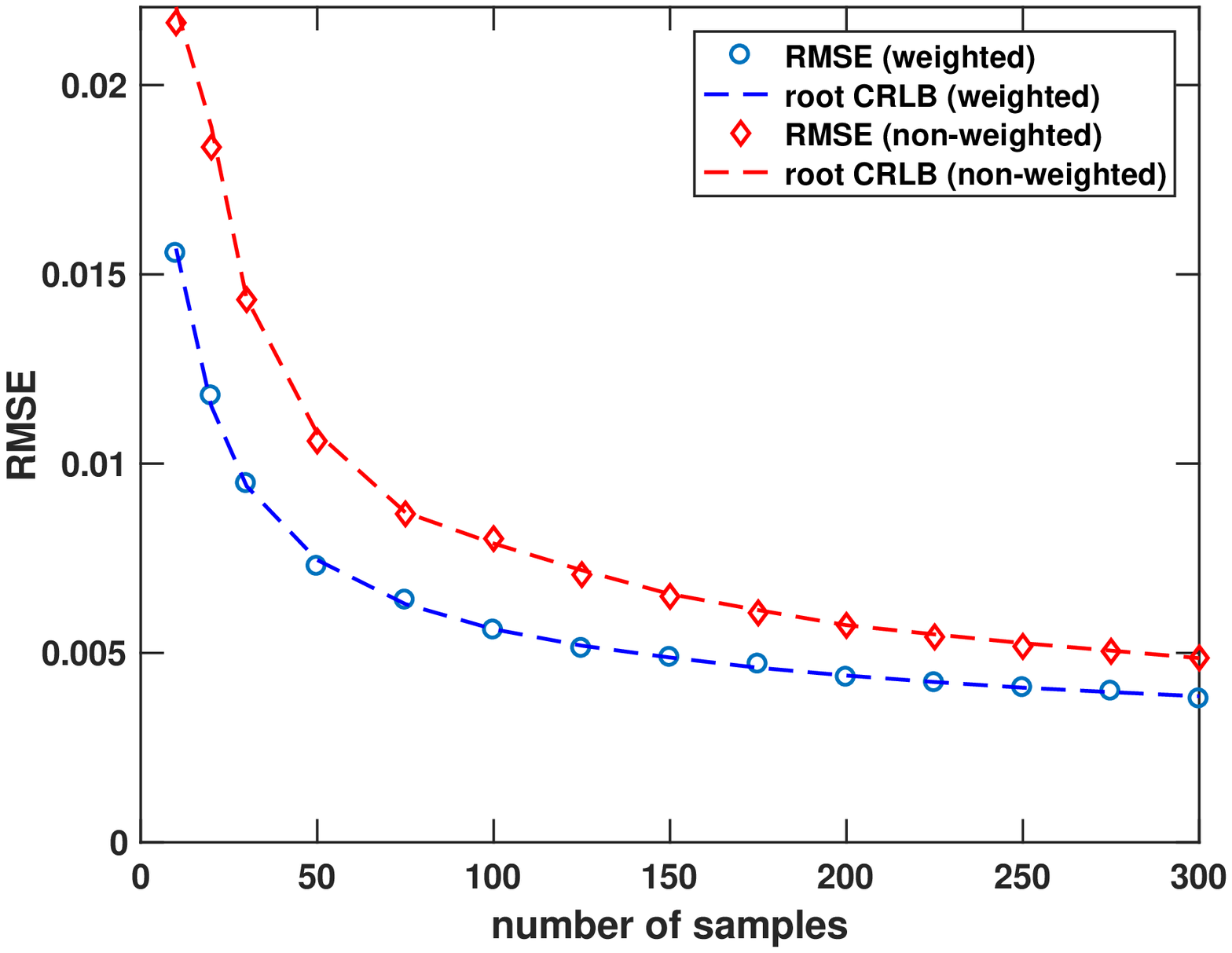}
           \caption{Obtained RMSE for the dampings in the first dimension, when using the sampling patterns for the weighted and non-weighted cases, respectively.}
            \label{fig:2D_sum_of_dampings_first_dim}
\end{figure}

\begin{figure}[t!]
        \centering
        \vspace{-2.05 mm}
            \includegraphics[width=0.48\textwidth]{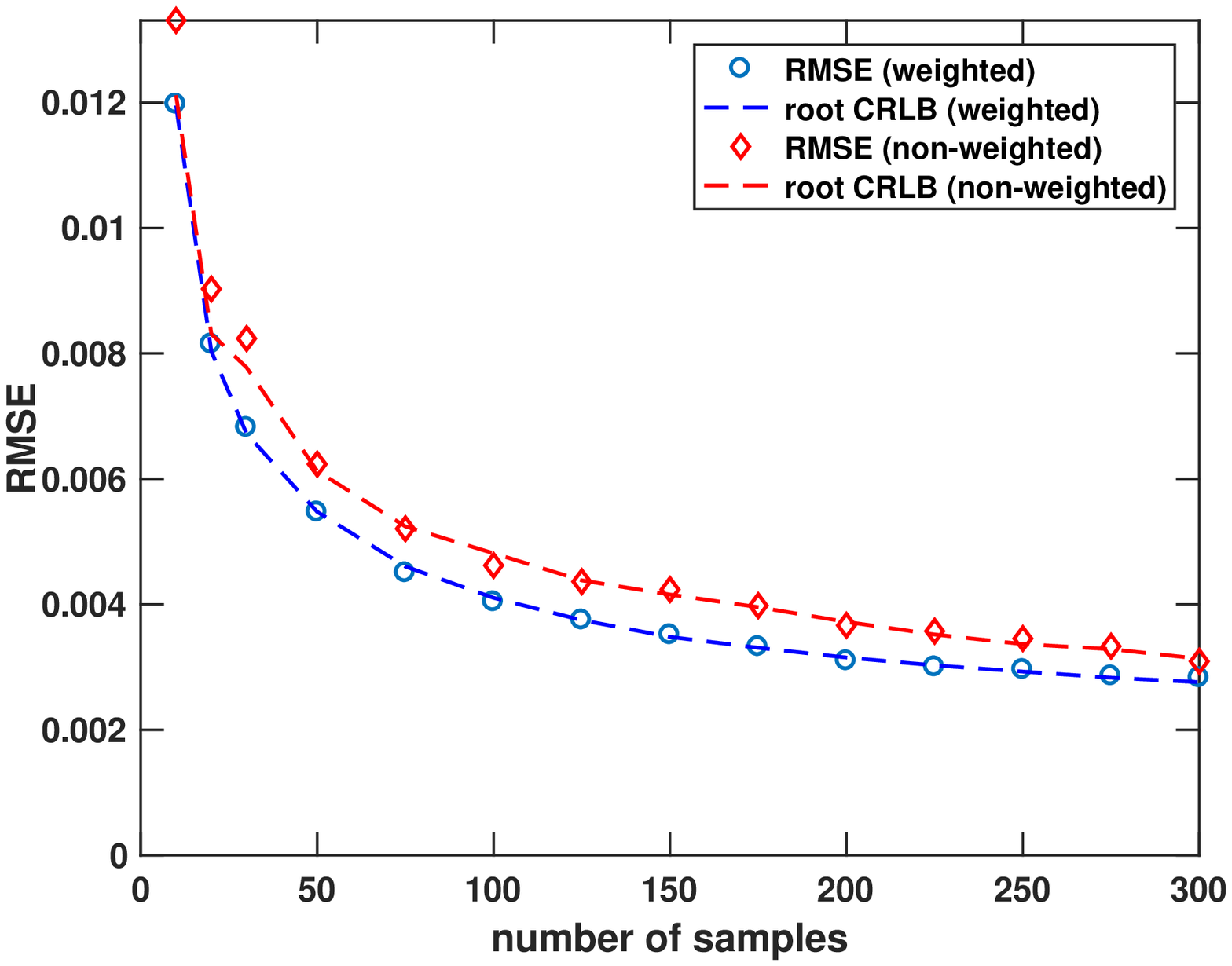}
           \caption{Obtained RMSE for the dampings in the second dimension, when using the sampling patterns for the weighted and non-weighted cases, respectively.}
            \label{fig:2D_sum_of_dampings_second_dim}
\end{figure}

\begin{figure}[t!]
        \centering
        \vspace{-2.05 mm}
            \includegraphics[width=0.48\textwidth]{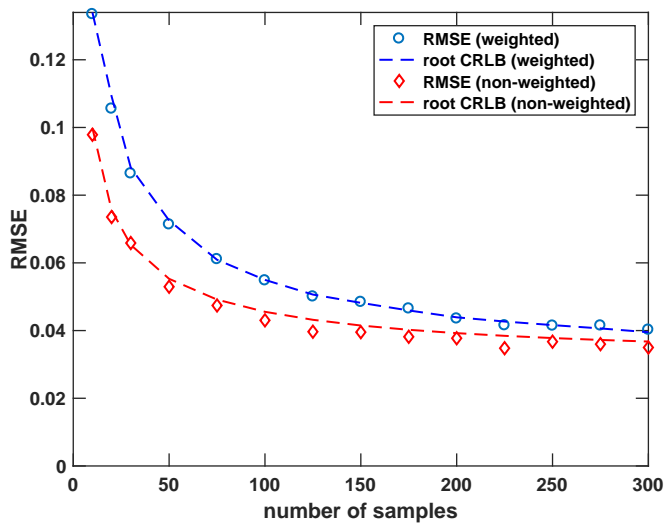}
           \caption{Obtained RMSE for the amplitudes, when using the sampling patterns for the weighted and non-weighted cases, respectively.}
            \label{fig:2D_sum_of_amps}
\end{figure}

We here consider the case of a signal consisting of two \mbox{2-D} damped sinusoid, i.e.,
\begin{align}
	y(t_1,t_2) = \sum_{k=1}^K\alpha_k\mathrm{e}^{i\phi_k} \Pi_{d=2}^2\mathrm{e}^{2i\pi f_{k,d}t_d - \beta_{k,d}t_d} + \epsilon(t_1,t_2)
\end{align}
for $K = 2$. Let the parameters be $(f_{1,1},f_{2,1}) = (0.1,0.2)$ and $(\beta_{1,1},\beta_{2,1}) = (0.1,0.1)$ for the first dimension, $(f_{1,1},f_{2,1}) = (0.1,0.2)$ and $(\beta_{1,1},\beta_{2,1}) = (0.1,0.1)$ for the second dimension, and let $\alpha_1 = 1$, $\alpha_2 = 1.3$, $\phi_1 = \frac{\pi}{3}$, $\phi_2 = \frac{\pi}{3}$, and $\sigma^2 = 0.01$. We then determine optimal sampling schemes by solving \eqref{eq:new_formulation} for varying number of sampling points. This is done for both the equally weighted case, i.e., with $\psi_p = 1$ for all $p$, as well as for the case when only the frequency and damping parameters are given weight, i.e., with $\psi_p = 0$ for the amplitude and phase parameters. The results are shown in Figures~\ref{fig:2D_sum_of_freqs_first_dim}-\ref{fig:2D_sum_of_dampings_second_dim}. In Figure~\ref{fig:2D_sum_of_freqs_first_dim}, the root of the sum of the CRLBs for the frequencies in the first dimension, i.e., $f_{1,1}$ and $f_{2,1}$, are shown. Similarly, Figure~\ref{fig:2D_sum_of_freqs_second_dim} corresponds to the frequencies in the second dimension, while Figures~\ref{fig:2D_sum_of_dampings_first_dim} and \ref{fig:2D_sum_of_dampings_second_dim} corresponds to the damping parameters in the first and second dimension, respectively. Also presented is the corresponding RMSE obtained when estimating the parameters using NLS. As can be seen, the obtained RMSEs coincides with the CRLBs for both the weighted and non-weighted case, implying that the bound is tight. Note also that the schemes corresponding to  assigning no weight to the amplitude and phase parameters all result in a lower sum of CRLB for the frequency and damping parameters than the non-weighted schemes. This comes at the price of a larger sum of CRLB for the amplitudes $\alpha_1$ and $\alpha_2$, which is illustrated in Figure~\ref{fig:2D_sum_of_amps}. As can be seen in the figure, the non-weighted sampling scheme here leads to more accurate estimates of the amplitudes.
%
%

\section{Conclusion}\label{sec:Conclusion}
In this work, we have proposed a convex optimization problem for finding suitable sampling schemes for multidimensional data models. The optimization problem is formed such that the number of used samples, chosen from a collection of available sampling points, is minimized while the sum of the variance of parameters of interest are guaranteed to be below a certain level. Due to the structure of the optimization problem, it is easy to add additional constraints, e.g., adding performance bounds on selected parameters, or putting more emphasize on a subset of the parameters, and to model for the uncertainty in \textit{a-priory} assumptions of the parameter values. In the numerical section, we show that solving the proposed optimization problem is a more efficient approach than randomly selecting the sampling points, especially in the multi-dimensional setting. Further, we show that using the sampling schemes found by solving the proposed optimization problem, will provide a lower Cram\'er-Rao lower bound than that found from using ordinary uniform sampling. By using an efficient parameter estimator on the signal sampled according to the found sampling scheme, we show that these Cram\'er-Rao lower bounds are, in fact, tight.
%

\bibliographystyle{IEEEbib}
\end{document}